\documentclass[twocolumn,showpacs,preprintnumbers,amsmath,amssymb,showkeys]{revtex4}

\usepackage{graphicx}
\usepackage{dcolumn}
\usepackage{bm}
\begin{document}

%\preprint{APS/123-SSP}
\title{Electronically Induced Anomaly in LO Phonon Dispersion\\ of High - $T_c$ Superconductors}

\author{Slaven Bari{\v s}i{\' c}}
\author{Ivana Mrkonji{\' c}}
 \email{ivanam@phy.hr}
\author{Ivan Kup{\v c}i{\' c }}
\affiliation{Department of Physics, University of Zagreb, POB 331,
10002 Zagreb, Croatia}

\date{\today}

\begin{abstract}

The strong, electronically induced anomaly in the spectrum
of the longitudinal optical (LO) phonons propagating along the main
axes of the CuO$_2$ plane is tentatively attributed to the oxygen-oxygen charge
transfer between the two oxygens in the plane.
    It is argued that this charge transfer can be large
and that it is strongly coupled to the zone boundary
LO phonons.
    The corresponding negative contribution to the free energy is quartic
in the LO phonon amplitude, making the LO phonon unstable through
the first order phase transition, with a concomitant domain
structure.
\end{abstract}



\maketitle

The discovery of the electron-induced anomaly in the dispersion
of the oxygen LO phonons, which
propagate along the main axes of the CuO$_2$ plane in
La$_{2-x}$Sr$_x$CuO$_4$ \cite{Queeney,Pinch1} and
YBa$_2$Cu$_3$O$_{6+x}$ \cite{Petrov},
has renewed interest in the origin of this anomaly and
its possible interplay with  high-$T_c$
superconductivity.
    Similar anomalies seem to occur in
La$_2$CaCuO$_{4+x}$ \cite{Pinch2},
Pb$_2$Sr$_2$(Ca,Y)Cu$_3$O$_8$ and
Li$_{1-x}$T$_{2-x}$O$_4$ \cite{Gompf}.
    The oxygen LO phonon
is anomalous even in the superconducting perovskite
Ba$_{1-x}$K$_x$BiO$_3$ \cite{Braden1,Braden2}.
    On the other hand, structural instabilities of a somewhat different
nature have been observed in La$_{2-x}$Ba$_x$CuO$_4$ \cite{Axe}
and Tl$_2$Ba$_2$CaCu$_2$O$_8$ \cite{Toby}, where the oxygen
displacements occur perpendicularly to the CuO$_2$ planes.

The common feature observed in  oxygen LO phonons of
La$_{2-x}$Sr$_x$CuO$_4$ (LSCO) and YBa$_2$Cu$_3$O$_{6+x}$ (YBCO)
is the occurrence of a break  in the dispersion at $2
\pi/a$(0.25,0), half-way between the origin (0,0) and the edge
(0.5,0) of the CuO$_2$ planar Brillouin zone (BZ). This break was
tentatively attributed \cite{Queeney} to the (quasi)static
longitudinal dimerisation of the oxygens with wave vector [0.5,0]
shown in Fig. 1a. The latter should produce the new BZ edge at
[0.25,0] and, correspondingly, a gap in the LO dispersion.
    However, the (quasi)static dimerisation has never been observed.
    This led to the speculation\cite{Queeney} that the dimerisation is highly disordered,
with various, usually very low estimates of the corresponding
correlation length.
    The length in question should not however be shorter than a few times $4a$,
if an LO phonon gap with  wavelength $4a$ is to be attributed to
the $2a$ dimerisation.

    \begin{figure}
    \includegraphics[height=10pc,width=20pc]{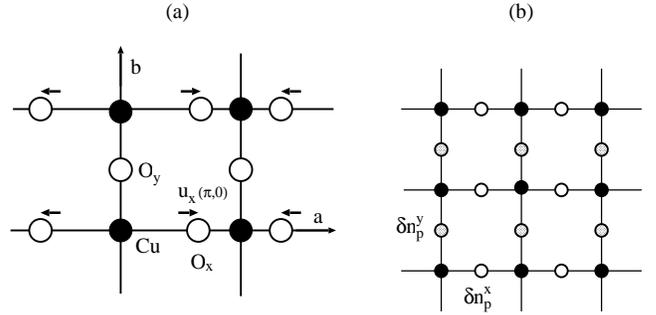}
    \caption{(a) The $(\pi, 0)$ oxygen LO phonon.
    (b)  The homogeneous component of the charge fluctuations
    induced by the $(\pi, 0)$ oxygen LO phonons.}
    \end{figure}

Although the LO phonon anomaly depends strongly on doping, the
electronic mechanism that produces it has not been understood
until now.
    The conventional Peierls-like explanation is unlikely,
because it requires the appearance of the strong [0.5,0] charge
density fluctuations, coupled to the lattice by a linear
electron-phonon coupling.
    Although some phonon induced features seem to
exist \cite{Lanzara}in the electron spectra, recent ARPES
measurements \cite{Shabel1,Shabel2,Ino1,Ino2} have however yielded
Fermi surfaces of LSCO and YBCO which do not possess the nesting
properties favoring  [0.5,0] charge fluctuations. On the other
hand, there is no apparent reason why the bare linear
electron-phonon coupling would itself be particularly strong
\cite{Kabanov} for the LO dimerisation.

The aim of this note is to examine a mechanism which can explain
the LO anomaly in LSCO and YBCO. When the linear electron-phonon
coupling is ruled out, the natural next step is to consider a
quadratic electron-phonon coupling and the charge fluctuations
which couple through it to the lattice. Such a quadratic
electron-phonon coupling was already used to explain the LTO/LTT
transition in La$_{2-x}$Ba$_x$CuO$_4$ \cite{SB}, in which the
oxygens move perpendicularly to the plane.
    When the LO dimerisation mode is involved quadratically in the electron-phonon coupling,
it can only couple to the homogeneous charge redistribution within
the unit cell. This charge redistribution can be either symmetric
or antisymmetric with respect to the symmetry operation $x
\rightleftharpoons y$. We will show here that the antisymmetric
charge redistribution is large, that it is exactly the same as the
one which leads to the LTO/LTT instability in
La$_{2-x}$Ba$_x$CuO$_4$, and that it explains satisfactorily the
LO instability in LSCO and YBCO.

The preceding symmetry analysis can be specified within the
tight-binding model, which distributes the holes over the Cu
$3d_{x^{2}-y^{2}}$ orbital and the O$_{x,y}$ $2p_{x,y}$ orbitals.
    This model is widely used in the theory of high-$T_{c}$ superconductors either in the
small $U$ or in the large $U$ limit, where $U$ is the Coulomb
repulsion on the Cu site. Only four single particle parameters are
usually used in addition to $U$, namely, the Cu and O$_{x,y}$ site
energies $\varepsilon_{d}$ and $\varepsilon^{x,y}_{p}$, the Cu-O
overlap $t$ and the O$_x$-O$_y$ overlap $t'$. \cite{IM}

The variation of the crystal-field potential of the ionic lattice,
introduced by the dimerisation modes with
the amplitudes $u_{\pi, 0}$ and $u_{0, \pi}$, leads
in the first place to the variation of the site energies,
i.e. to a variation of $\Delta_{pd}=\varepsilon_{d}+
(\varepsilon^{x}_{p}+\varepsilon^{y}_{p})/2$ and
$\Delta_{pp}=\varepsilon^{x}_{p}-\varepsilon^{y}_{p}$.
    The variation of the oxygen site energies can be
calculated in the point-charge approximation, and written
in the form \cite{Batistic,Kupcic}
\begin{equation}
\label{1}
\varepsilon^{i}_{p} \approx \frac{2e^2}{a} [\alpha_p^i
u_{Q} + \beta_p^i u_{Q}^2 + \cdots],
\end{equation}
where $Q =$ $(\pi, 0)$ or $(0,\pi)$. Previous calculations
\cite{Kupcic} estimated $\alpha_p^i$ and $\beta_p^i$ from the
crystal field potential at the oxygen sites. The values of
$\alpha_p^i$ and $\beta_p^i$ obtained in this way for LSCO and
YBCO are given in Table 1.
\begin{table}[h]
\caption{(a) The electron-phonon coupling constants
$\alpha_p^i$ and $\beta_p^i$
for the $(\pi, 0) $ deformation.
    (b) The corresponding  constants calculated for
the LTT deformation of the La$_{2}$CuO$_4$ lattice.
}
   \begin{center}
     \begin{tabular}{cccccc} \hline
 & & $\alpha_p^x$ & $\beta_p^x$ & $\alpha_p^y$& $\beta_p^y$ \\ \hline
(a) &YBa$_2$Cu$_3$O$_{7}$ & 0.0 & -0.2 & 1.5  & 0.9  \\
 &La$_{1.8}$Sr$_{0.2}$CuO$_4$  & 0.05 & 1.1 & 2.1 & 0.9  \\ \hline
(b) &LTT  & 0.0 & -0.9 & 0.0 & -1.8  \\ \hline
     \end{tabular}
   \end{center}
 \end{table}

Singling out the homogeneous component of $\Delta_{pp}$,
it is found that

\begin{eqnarray}
\delta \Delta^{H}_{pp} & \approx & (\beta_p^x - \beta_p^y)
(u^{2}_{\pi,0}-u^{2}_{0,\pi}),
\nonumber \\
\label{2} \delta \Delta^{H}_{pd}& \approx & \frac{1}{2} (\beta_p^x
+ \beta_p^y)
 (u^{2}_{\pi,0}+u^{2}_{0,\pi}).
\end{eqnarray}

    $\delta \Delta^{H}_{pp}$ couples to the homogeneous charge
redistribution between O$_x$ and O$_y$,
$n^{H}_{pp}=(n^{x}_{p}-n^{y}_{p})^{H}$ shown in Fig. 1b,
adding a bilinear term to the Hamiltonian
$H_{0}$,
\begin{equation}
\label{3}
 H=H_{0}+n^{H}_{pp} \delta \Delta^{H}_{pp}.
\end{equation}

    In most of the high-$T_c$ superconductors(YBCO is an exception) $\Delta^{H}_{pp}=0$
in the undeformed lattice and $\delta \Delta^{H}_{pp}$ lifts the
$\varepsilon^{x}_{p}=\varepsilon^{y}_{p}$ degeneracy of the
CuO$_2$ unit cell, in a way similar to the Jahn-Teller (JT)
effect.
    Even in the presence of the
hybridisation $t$ and $t'$, the $\delta \Delta^{H}_{pp}$
JT effect leads to  large energetic gains.
    In the conventional JT effect the charge transfer
between the two levels,
whose  degeneracy is lifted, is a
step function of the splitting $\delta \Delta^{H}_{pp}$.
    In general, the effect of hybridisations $t$ and $t'$
is however to make this charge transfer analytic
\begin{equation}
\label{4}
\bar{n}^{H}_{pp} \approx \chi_{pp} \delta
\Delta^{H}_{pp},
\end{equation}
where $\chi_{pp}$ is the exact charge susceptibility for the
Hamiltonian (\ref{3}), i.e. $\chi_{pp}$ is related to the exact
correlation function $\langle n^{H}_{pp}n^{H}_{pp} \rangle$ for
the Hamiltonian $H_0$ of  (\ref{3}).
    The JT effect is now associated with the expectation
that $\chi_{pp}$ is large.
    Actually, a singular $\chi_{pp}$ in  (\ref{4})
indicates a nonanalytic nature of the charge transfer $\bar{n}^{H}_{pp}$,
i.e. recovery of the conventional JT effect, with $\Delta F$ linear
in $\Delta^{H}_{pp}$.

The free energy variation, corresponding to  (\ref{4}), is
\begin{equation}
\label{5}
 \Delta F=- \frac{1}{2} \chi_{pp} \delta
{\Delta^{H}_{pp}}^{2}+ \cdots .
\end{equation}
    It should be noted that, according to  Refs.(\ref{2}) and (\ref{4}),
$\Delta F$ of the equation (\ref{5}) is quartic in the
dimerisation $u_{\pi, 0}$.
    However, when $\chi_{pp}$ is singular,
$\Delta F$ is to be taken linear in $\delta \Delta _{pp}$, i.e. quadratic
in dimerisation.

Large $\chi_{pp}$ is found every time when the Fermi level falls
close to the two groups of the quasi degenerate electronic states
respectively associated with the $a$ and $b$ axes of the CuO$_2$
plane, which get split by the finite $\delta \Delta^{H}_{pp}$.
    This is in particular the case when the Fermi level falls close to the van Hove singularities
at (0.5,0) and (0,0.5) points of the two-dimensional Brillouin
zone, associated with the single electron propagation in the
CuO$_2$ planes.
    The single particle picture can be used with some confidence
in two limits for the Hamiltonian $H_0$ in (\ref{3}), namely for
free fermions $(U=0)$ and for infinite $U$ ($U$ larger than
$\Delta_{pd}, \Delta_{pp}, t$ and $ t'$).
    In this case, for $T \approx 0$ K, the real part of the generalized susceptibility,
given in terms of the vertex functions $g({\bf k})$ and $ h ({\bf k})$,
reads as
 \begin{eqnarray}
 \label{6}
{\rm Re} \{ \chi _{g,h} (\varepsilon_F) \}&=&
 \frac{1}{V} \sum_{{\bf k} \sigma} g({\bf k}) h({\bf k})
\frac{ \partial f [\varepsilon_L ({\bf k})]}{
 \partial \varepsilon_L ({\bf k})}.
\end{eqnarray}
    Here $f [\varepsilon_L ({\bf k})]$ is the Fermi function and
$\varepsilon_L ({\bf k})$ is the energy of the lowest occupied
band \cite{IM}.
    The case $g({\bf k}) = h ({\bf k}) = 1$ describes
the density of states at the Fermi level, $n_F$,
averaged appropriately over $k_B T$,
and $g({\bf k}) = h ({\bf k}) =
\partial \varepsilon_L ({\bf k})/ \partial \Delta_{pp}$
the real part of the intraband $pp$ susceptibility, $\chi_{pp} $.
    Although it has been shown
\cite{IM} that interband effects might also be important in
high-$T_c$ materials, for the sake of simplicity we shall retain
here only  the intraband contribution to $\chi_{pp} $.
    Assuming that $\Delta_{pp}^H = 0$ at the outset and
for $\varepsilon_{F} \approx \varepsilon_{vH}$,
it follows than straightforwardly that
\begin{equation}
\label{7}
 - \chi_{pp}  \approx
 c\cdot n_{F} \approx  c \cdot \log \frac{\varepsilon_{0}}{{\rm max}[k_{B}T,
|\varepsilon_{F}-\varepsilon_{vH}|]},
\end{equation}
with the constant $c$ typically of order $1/5$, as illustrated in Fig. 2.
    $\varepsilon_0$ is the energy scale which characterizes
the van Hove singularity at $\varepsilon_{vH}$,
assumed two-dimensional, logarithmic, rather than extended one-dimensional,
i.e. more singular than the logarithm.
    $\chi_{pp}$ of Eq. (\ref{7}) is singular at low temperatures for
$\varepsilon_{F} \approx \varepsilon_{vH}$, when the expansion
(\ref{5}) breaks down and $\Delta F$ becomes linear in $\delta
\Delta^{H}_{pp}$, i.e. quadratic in the dimerisation displacement
$u$.
    This reasoning can be applied to YBCO by the replacement
$\delta \Delta_{pp}^H \rightarrow \Delta_{pp}^H + \delta
\Delta_{pp}^H$ in Refs.(\ref{4}) and (\ref{5}), assuming that the
value of $\Delta_{pp}^H$ in the undeformed lattice is sufficiently
small.

\begin{center}
    \begin{figure}[htb]
    \includegraphics[height=15pc,width=15pc]{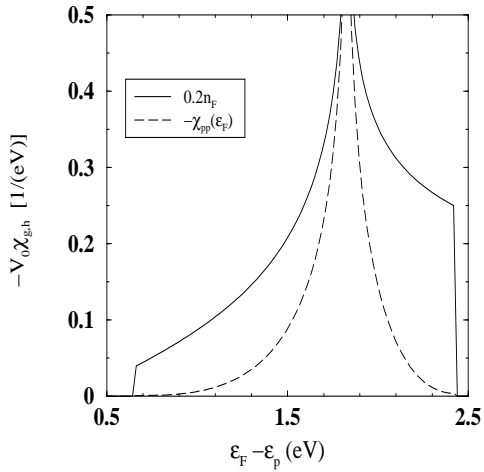}
    \caption{The density of states
    and the  intraband $pp$ susceptibility as functions of the Fermi
    energy $\varepsilon_F$, for the simplest case,
    $\Delta_{pd} = 0.66$ eV, $t = 0.73$ eV,
    and $t' = \Delta_{pp} = U = 0$.
    }
    \end{figure}
\end{center}

The LO anomaly is observed at a finite temperatures when $n_F$ of
(\ref{3}) is finite, i.e. the negative contribution to the free
energy $\Delta F$ can be taken as quartic in $u$. As discussed
previously \cite{SB}, the activation of the negative contribution
to $\Delta F$ quartic in $u$ occurs through first order phase
transition. The compelling feature of such explanation is that the
two phases with $u=0$ and a finite $u$ mix, i.e. the (disordered)
domains of the dimerised phase are expected to condense in the
$u=0$ phase. This might explain the difficulty in the experimental
observation of the dimerised domains, provided that the latter
turns out to be small. The theoretical discussion of the domain
size requires the insight into the gradient terms quadratic and/or
perhaps quartic in $u$, which is however beyond the scope of the
present paper.

Here, we only wanted to stress that the homogeneous charge
transfer couples appreciably to the LO dimerisation because the
electron-phonon coupling constants of Table I are large and
because the corresponding susceptibility (\ref{7}) is large. The
disadvantage that the electron-phonon coupling is quadratic rather
than linear in $u$ is set off by satisfactory consequence that the
resulting phase transition is of  first order, providing a
possible explanation of the LO anomaly observed in LSCO and YBCO.
    Further experimental and theoretical investigations are
however required to explain unambiguosly the origin of the LO
anomaly in the high-$T_c$ superconductors.\\

\begin{acknowledgments}

Useful discussions with Lev Gor'kov, Jacques Friedel and Robert
Comes are gratefully acknowledged.
\end{acknowledgments}

\end{document}